\documentclass[10pt,aps,pra,twocolumn,superscriptaddress,floatfix,longbibliography]{revtex4-2}
\usepackage{placeins}
\usepackage[english]{babel}
\usepackage{xcolor}
\usepackage[utf8]{inputenc}
\usepackage{mathtools}
\usepackage{physics}
\usepackage{braket}
\usepackage{bbm}
\usepackage{xspace}
\usepackage{bm}
\usepackage{amsbsy}
\usepackage{amsthm}
\usepackage{amssymb}
\usepackage{lipsum}
\usepackage{amsfonts}
\usepackage{amsmath}
\usepackage{soul}
\setstcolor{red}
\usepackage{dsfont}
\usepackage{graphicx}
\usepackage{float}
\usepackage{hhline}
\usepackage{epsfig}
\usepackage{epstopdf}
\usepackage{dsfont}
\usepackage{xcolor}
\usepackage{xcolor}
\usepackage{color}
\usepackage{verbatim}
\usepackage{nomencl}
\usepackage[normalem]{ulem}
\usepackage{multirow}
\usepackage{makecell}
\usepackage{url}
\usepackage{tikz}
\usetikzlibrary{arrows.meta, decorations.pathreplacing}
\DeclareUnicodeCharacter{2215}{/}
\DeclareUnicodeCharacter{0301}{\'{e}}
\usepackage[colorlinks=true,linkcolor=blue,citecolor=blue,urlcolor=blue]{hyperref}

\newcommand{\blue}[1]{\textcolor{blue}{#1}}

\begin{document}

\title{Majorana bound states in a hybrid Kitaev ladder with long-range pairing}
\author{Rajiv Kumar}
\email{rajivkumar.rs.phy22@iitbhu.ac.in}
\affiliation{Department of Physics, \blue{Indian Institute of Technology (Banaras Hindu University)} Varanasi - 221005, India}
\author{Tapan Mishra}
\email{mishratapan@niser.ac.in}
\affiliation{School of Physical Sciences, \blue{National Institute of Science Education and Research,} Jatni 752050, India}
\affiliation{\blue{Homi Bhabha National Institute,} Training School Complex, Anushaktinagar, Mumbai 400094, India}
\author{Levan Chotorlishvili}
\email{levan.chotorlishvili@gmail.com}
\affiliation{Department of Physics and Medical Engineering, \blue{Rzeszow University of Technology,} 35-959 Rzeszow Poland}
\author{Sunil Kumar Mishra}
\email{sunilkm.app@iitbhu.ac.in}
\affiliation{Department of Physics, \blue{Indian Institute of Technology (Banaras Hindu University)} Varanasi - 221005, India}

\begin{abstract}
We investigate an inter-leg coupled hybrid Kitaev ladder composed of two parallel superconducting chains with distinct pairing interactions. The upper chain of the ladder hosts conventional  $p$-wave pairing, while the lower chain exhibits long-range pairing that decays algebraically with distance. We demonstrate that the mutual influence of long-range pairing exponent, chemical potential, and inter-leg coupling strength gives rise to a rich topological phase diagram characterized by multiple Majorana zero modes and massive Dirac modes. In particular, we show that the inter-leg coupling renormalizes the effective energy scales, leading to a systematic shift of the topological phase boundaries and enabling controlled tuning of the Majorana modes. Furthermore, we identify a transition from a two Majorana zero mode phase to a phase encapsulating four Majorana zero modes, as the long-range pairing exponent is varied. This transition is accompanied by a crossover regime in which Majorana zero modes coexist with massive Dirac modes, reflecting hybridization between edge and bulk excitations. This ladder thus provides a minimal and attractive platform for realizing the impact of a long-range pairing on topological phases. Our results highlight the potential of long-range hybrid systems for engineering tunable topological states relevant for quantum information applications.
\end{abstract}

\maketitle

\section{Introduction}

Majorana edge modes in topological superconductors have been the subject of advanced research in condensed matter physics over the past decade \cite{A_Y_Kitaev-2001, PhysRevLett.104.040502, PhysRevLett.105.177002}. These quasiparticle excitations, mainly Majorana zero modes (MZMs) localized at edges of chain, obey non-Abelian exchange statistics and exhibit intrinsic robustness against local perturbations. Due to these properties, MZMs are widely considered as a promising candidate for fault-tolerant topological quantum computation \cite{RevModPhys.80.1083, KITAEV20032}. The conventional Kitaev chain and the long-range (LR) Kitaev chain  constitute a prototype model for the emergence and manipulation of  MZMs and Massive Dirac modes (MDMs) localized at the edges \cite{PhysRevResearch.6.033154, PhysRevB.107.035440}. Recent advances in engineered quantum platforms have enabled the realization of Kitaev models and their extended versions, allowing controlled access to topological properties and facilitating experimental studies of edge mode physics \cite{PhysRevB.88.020407, PhysRevB.105.125418, PhysRevB.104.235423}.  
One possible way of realizing the Kitaev chain is the proximity of semiconductor nanowires coupled to superconductors in the presence of strong Rashba spin-orbit interaction \cite{6qwr-4679, PhysRevLett.105.077001, Nehra_2020}, as well as in planar Josephson junction architectures \cite{PhysRevLett.118.107701, PhysRevB.111.195433, PhysRevB.97.235114}. Substantial experimental progress on such platforms has enabled controlled investigations of MZMs and related topological behaviours \cite{PhysRevB.109.035415,  r9pv-2prs, Zatelli_2024}. These efforts have progressed through the potential application of MZMs to topological computation and through the accumulation of experimental indications consistent with their emergence in topological systems. However, the unambiguous identification of MZMs remains a significant challenge, such as disorder \cite{PhysRevB.103.224505, q488-97dp, 2v41-yvs1}, impurities \cite{c7lm-323v, PhysRevB.110.214506, bksr-fqn6}, and the presence of quasi-MZMs can produce experimental signatures in a minimal Kitaev chain \cite{PhysRevB.61.10267, hszk-ymf3}. These complications make it difficult to distinguish genuine topological states from trivial counterparts in realistic systems. Within the minimal Kitaev chain, MZMs can also appear under finely tuned conditions in parameter space, often referred to as “sweet spots” \cite{PhysRevB.109.035415, 7367-wbp9, klwd-xckq}. However, such states do not possess robust topological protection and are highly sensitive to perturbations. For this reason, they are frequently described as “poor man’s Majorana modes”, reflecting their lack of stability compared to true MZMs \cite{PhysRevB.86.134528, PhysRevB.110.L180402, PhysRevB.110.125408}.
\par
Although the prototype model, the Kitaev chain is identified by hopping and pairing of nearest neighbours (NN), it has been shown that extending the superconducting pairing beyond the NN \cite{Maity_2020, PhysRevLett.118.267002, PhysRevB.95.195160, PhysRevLett.113.156402, PhysRevLett.119.110601, PhysRevB.111.104308, PhysRevB.110.064302, PhysRevB.104.075113, PhysRevB.94.125121, PhysRevB.96.125113, PhysRevB.97.115436, PhysRevB.98.134507, PhysRevB.110.064302, Yan_2023, Bhattacharya_2019}, or hopping and superconducting pairing beyond the NN \cite{PhysRevLett.118.267002, PhysRevB.95.195160} can qualitatively modify their topological properties. In particular, LR pairing term that includes algebraic decay exponent $\alpha$ leads to anomalous topological phases and unconventional Majorana excitations. Depending on the exponent $\alpha<1$, LR Kitaev chains host MDMs excitations per edge or exhibit nonlocal edge states, and for $\alpha>1$, there exist MZMs per edge. The LR Kitaev chain has recently been introduced to describe systems such as helical Shiba chains, which arise from arrays of magnetic impurities deposited on the surface of a s-wave superconductor \cite{PhysRevB.88.155420, science_1259327, Kim2020, PhysRevB.106.165428, PhysRevApplied.12.034048}.  Beyond the extension of the LR Kitaev chain, such as coupled and hybrid Kitaev models that include distinct pairing interactions, can host multiple Majorana bound states (MBSs) [including MZMs and MDMs] localized at the edges of the systems \cite{7cqp-ws6c, n7bl-slgm, f56n-l1yr}. Recent studies of a Kitaev chain and their ladder have explored the topological properties and behaviour of MBSs \cite{Almeida_2021, PhysRevB.98.024205, PhysRevB.106.165428, PhysRevResearch.4.033088, Maiellaro_2019, Maiellaro_2018, PhysRevB.104.075113, r9zk-5wny, f56n-l1yr, k4gw-7k8h, PhysRevResearch.2.013175, PhysRevB.99.174303, PhysRevB.89.174514, PhysRevA.99.043624, PhysRevResearch.4.033088, PhysRevB.93.165142, PhysRevResearch.7.023183, hxb3-g7pl, PhysRevB.93.155425, PhysRevB.96.184516, maiellaro2018topological, maiellaro2019unveiling, f1m4-vkvq, PhysRevA.110.022212, PhysRevB.107.125110}. Several experiments have given evidence of MBSs \cite{PhysRevLett.110.126406, PhysRevLett.110.217005}.
\par
\begin{figure}[!t]
\begin{tikzpicture}[scale=1.7,>=Stealth,
    site/.style={circle,draw=brown,very thick,minimum size=10mm,inner sep=0pt},
    top/.style={site,fill=orange!25},
    bot/.style={site,fill=yellow!25},
    nn/.style={line width=1.5pt},
    vert/.style={line width=1.5pt,dashed,draw=teal!80!black},
    lr/.style={line width=1.5pt,densely dotted,draw=red!70!black}
]

\def\dx{1.0}
\def\dy{1.5}

\node[top] (A1) at (0,0) {\Large{$c_1$}};
\node[top] (A2) at (\dx,0) {\Large{$c_2$}};
\node[top] (A3) at (2*\dx,0) {\Large{$c_3$}};
\node at (2.65*\dx,0) {$\cdots \cdots \cdots$};
\node[top] (ANm1) at (3*\dx,0) {\Large{$c_{L-1}$}};
\node[top] (AN)   at (4*\dx,0) {\Large{$c_L$}};
\node[bot] (B1) at (0,-\dy) {\Large{$d_1$}};
\node[bot] (B2) at (\dx,-\dy) {\Large{$d_2$}};
\node[bot] (B3) at (2*\dx,-\dy) {\Large{$d_3$}};
\node at (2.65*\dx,-\dy) {$\cdots \cdots \cdots$};
\node[bot] (BNm1) at (3*\dx,-\dy) {\Large{$d_{L-1}$}};
\node[bot] (BN)   at (4*\dx,-\dy) {\Large{$d_L$}};

\draw[nn] (A1)--(A2) node[midway,above=3pt] {\large{$t,\Delta$}};
\draw[nn] (A2)--(A3) node[midway,above=3pt] {\large{$t,\Delta$}};
\draw[nn] (ANm1)--(AN) node[midway,above=3pt] {\large{$t,\Delta$}};
\draw[nn] (B1)--(B2) node[midway,below=3pt] {\large{$t,\Delta$}};
\draw[nn] (B2)--(B3) node[midway,below=3pt] {\large{$t,\Delta$}};
\draw[nn] (BNm1)--(BN) node[midway,below=3pt] {\large{$t,\Delta$}};
\draw[lr] (B1) to[bend left=15] (B3);
\draw[lr] (B1) to[bend left=15] (BNm1);
\draw[lr] (B1) to[bend left=15] (BN);
\draw[vert] (A1)--(B1) node[midway,left=3pt] {\Large{$t_v$}};
\draw[vert] (A2)--(B2) node[midway,left=3pt] {\Large{$t_v$}};
\draw[vert] (A3)--(B3) node[midway,left=3pt] {\Large{$t_v$}};
\draw[vert] (ANm1)--(BNm1) node[midway,left=3pt] {\Large{$t_v$}};
\draw[vert] (AN)--(BN) node[midway,left=3pt] {\Large{$t_v$}};

\end{tikzpicture}
\caption{Schematics of the inter-leg coupled hybrid Kitaev ladder. The upper leg composed of sites $c_1,c_2,\ldots,c_L$, hosts NN pairing, while the lower leg consisting of sites $d_1,d_2,\ldots,d_L$, supports LR pairing represented by dotted arcs. The two chains are coupled through vertical inter-leg tunneling $t_{\nu}$ between corresponding lattice sites $(c_i,d_i)$.}
\label{fig_kitaev_ladder}
\end{figure}
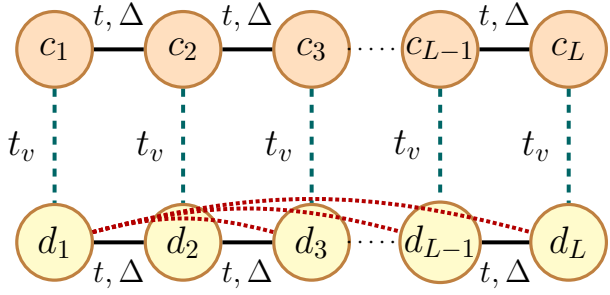

In this work, we investigate an inter-leg coupled hybrid Kitaev ladder consisting of a conventional Kitaev chain coupled in parallel to a LR Kitaev chain. These two chains are locally connected via inter-leg coupling, forming an asymmetric ladder geometry in which NN interaction and LR superconducting pairing coexist, thereby realizing a \textit{hybrid Kitaev ladder} as shown in Fig.~\ref{fig_kitaev_ladder}. The resulting hybrid Kitaev ladder provides a minimal platform, which is composed of NN and LR pairing. In particular, we focus on how the LR pairing in one leg modifies the bulk-boundary correspondence and influences the behaviour of MZMs and MDMs for fixed inter-leg coupling strength in the ladder. Further, we analyze the hybridization of MZMs and MDMs across the hybrid Kitaev ladder and demonstrate the emergence of topological proximity effect induced by inter-leg coupling and LR exponent. This coupling also renormalizes the effective energy scales of the system, leading to a shift of the topological phase boundaries and provides a controlled route for tuning the number and stability of edge modes. The hybrid Kitaev ladder thus serves as a versatile setting for exploring how anomalous topological phases influence Majorana physics in low-dimensional topological systems. Overall, our results suggest that the interplay of inter-leg coupling, the LR exponent, and engineered ladder geometries provide a powerful route for designing controllable and robust platforms for topological quantum computation.

\par
The present study is organized as follows. In Sec.~\ref{Model}, we formulate the hybrid Kitaev ladder and analyze  energy spectrum, with particular emphasis on the effects of inter-leg coupling and the LR exponent. Sec.~\ref{Symmetry_DR} is devoted to the symmetry analysis and dispersion relations. In Sec.~\ref{PD_HKL}, we analyze the phase diagram of the ladder. Finally, Sec.~\ref{conclusion} summarizes the results and possible directions for future research.
\begin{figure}
\includegraphics[width=0.48\linewidth,height=0.5\linewidth]{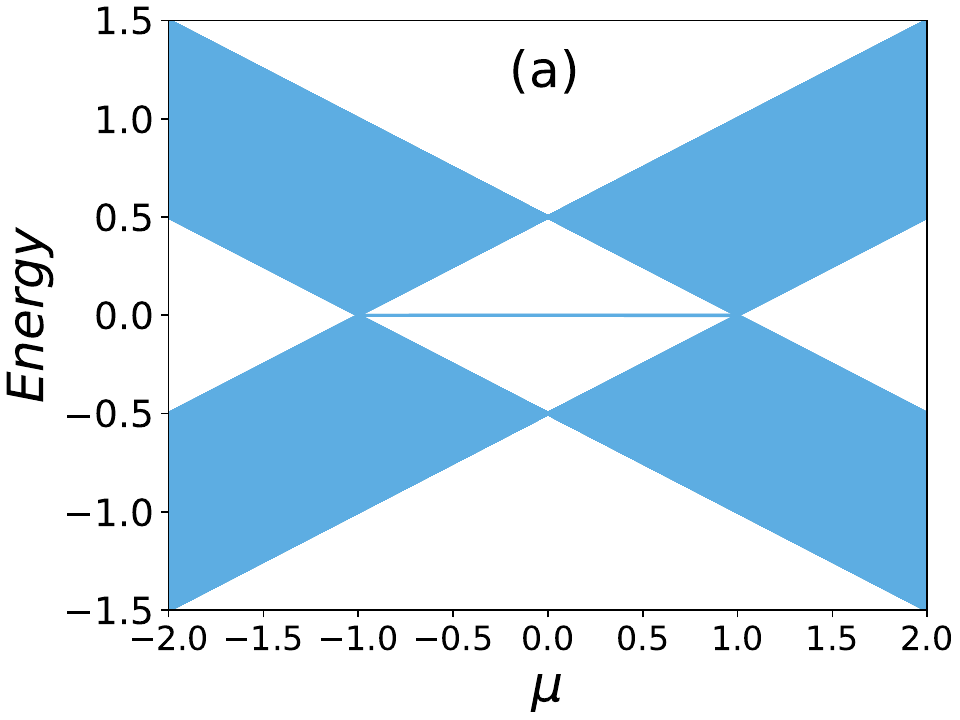}
\includegraphics[width=0.48\linewidth,height=0.5\linewidth]{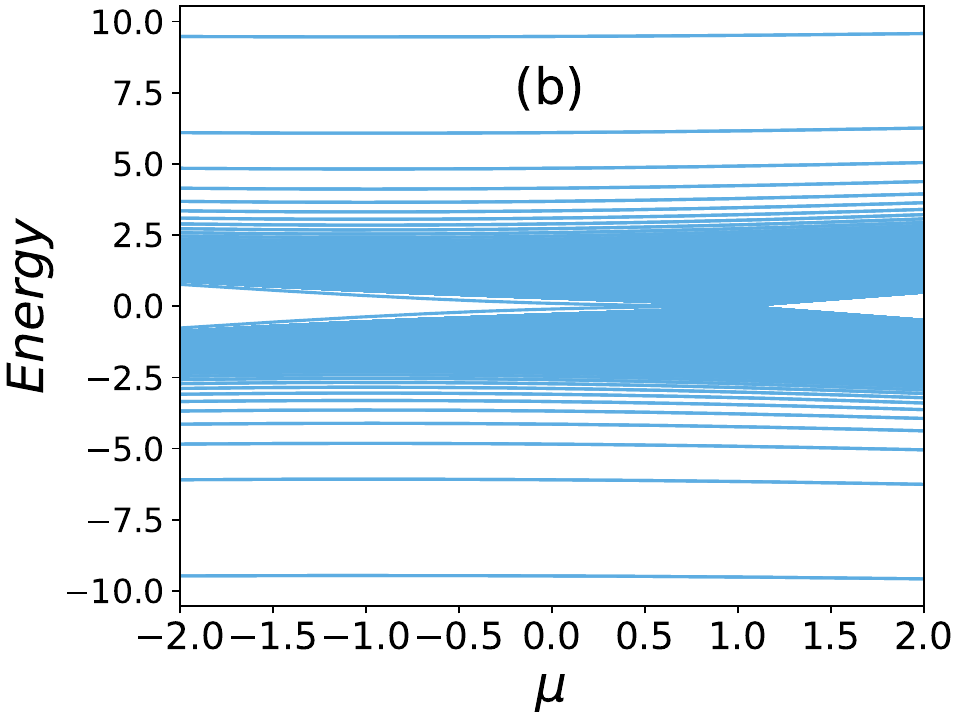}
\includegraphics[width=0.7\linewidth,height=0.5\linewidth]{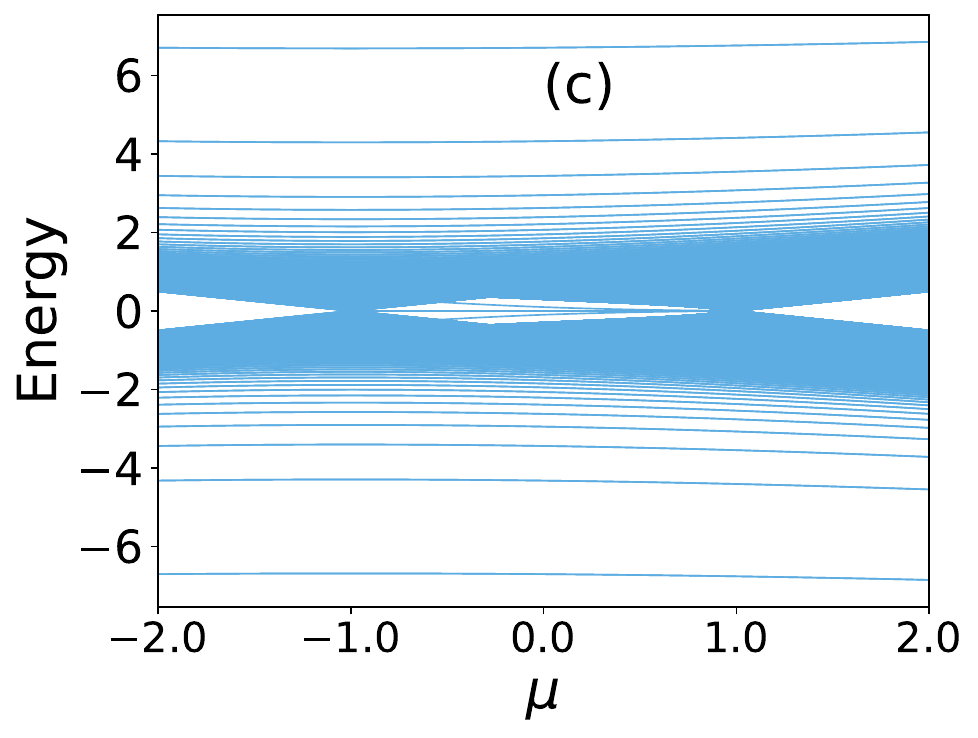}
\caption{Energy spectrum for the (a) conventional Kitaev chain and (b) LR Kitaev chain (c) Hybrid Kitaev ladder with $t_\nu = 0.0$ under OBC. The parameters used are $t=\Delta=1.0$ and $L=1000$, with $\alpha=0.5$ for the LR pairing shown in panel (b).}
\label{EG_KC_LKC}
\end{figure}

\section{Model}
\label{Model}
\subsection*{A. Long-range Kitaev chain}

We consider a one-dimensional LR Kitaev chain of $L$ spinless fermions subject to the open boundary condition (OBC). The Hamiltoian of the system is given by
\begin{equation}
\begin{aligned}
H_{LR} = -t  \sum_{i=1}^{L-1} & \left( c_{i}^{\dagger}c_{i+1} + H.c.\right) - \mu \sum_{i=1}^{L}  \left(2c_{i}^{\dagger}c_{i}-1\right) \\
& + \Delta \sum_{i=1}^{L-1} \sum_{l=1}^{L-j} \frac{1}{\delta_{l}^{\alpha}}  \left( c_{i}c_{i+l} + H.c.\right).
\label{LKC}
\end{aligned}
\end{equation}

In this Hamiltonian, the operators $c_i$($c_i^\dagger$) annihilate (create) a spinless fermion at the site $i$, obeying the standard fermionic anti-commutation relations $\{c_i, c_j^\dagger\} = \delta_{i,j}$ and $ \{c_i^\dagger, c_j^\dagger\} =\{c_i, c_j\} = 0 $. The parameters $t,~\Delta,~ \mu$ and $\alpha$ represent the hopping amplitude, the superconducting pairing strength, the chemical potential and the LR pairing exponent, respectively. In this work, we set the parameters $t = \Delta = 1$ and $\alpha=0.5$. The LR pairing, with an algebraic decay \(1/\delta_l^{\alpha}=1/|i-j|^{\alpha}\) for \(i<j\) is characterized by the exponent $\alpha$: the smaller values $\alpha<1$ lead to a slow algebraic decay with distance, referring to true LR, while the limit $\alpha \to \infty$ recovers the Kitaev chain \cite{Maity_2020, PhysRevB.105.085106}. The energy spectra are shown for the Kitaev chain in Fig.~\ref{EG_KC_LKC}(a) and for the LR Kitaev chain in Fig.~\ref{EG_KC_LKC}(b). The energy spectrum Fig.~\ref{EG_KC_LKC}(a) exhibits particle-hole (PH), time reversal (TR) and chiral symmetry, with a topological phase regime occurring within the transition points $\mu = \pm t$, with MZMs at the edges. For $|\mu| > t$, the spectrum reveals a trivial phase.
\par
The energy spectrum for the LR Kitaev chain with exponent $\alpha<1$ exhibits substantial modifications, as shown in Fig.~\ref{EG_KC_LKC}(b). In this regime, the chain supports a topological phase for $\mu<t$, with MDMs, they localize at the edges, while the system becomes topologically trivial for $\mu>t$. These MDMs originate from the hybridization of MZMs as a result of the LR pairing. Moreover, both the localization properties and the excitation gap of the MDMs are strongly influenced by $\alpha$ and $\mu$. The qualitative differences between the conventional Kitaev chain and the LR Kitaev chain provide a natural starting point for investigating the hybrid Kitaev ladder. In the following, we consider two distinct situations: the hybrid ladder formed by coupling a NN Kitaev chain with an LR Kitaev chain, and the limiting case $\alpha>1$, where the system effectively approaches a NN-NN Kitaev ladder. This framework enables us to systematically explore how the inter-leg coupling and LR exponent modify the energy spectrum, reshape the topological phase diagram, and control the emergence of edge modes.
\subsection*{B. Hybrid Kitaev Ladder}
\label{HKL}
In this section, we study a hybrid Kitaev ladder composed of the Kitaev chain coupled in parallel to the LR Kitaev chain shown in Fig.~\ref{fig_kitaev_ladder}, described by fermionic operators \(c_i\) and \(d_i\) with their conjugates in the upper and lower chains, respectively. The inter-leg coupling strength \(t_{\nu}\) is present between the corresponding chain site \((c_i,~d_i)\). The upper chain \((c_i)\) hosts only NN interactions, while the lower chain (\(d_i\)) hosts LR pairing interactions same as before, while the terms of hopping and chemical potential remain strictly local. This asymmetric superconducting system enables a controlled investigation of the Majorana modes hybridization and emergent topological behaviour.
\par
The ladder Hamiltonian is the expressed sum of the four terms: the hopping terms, the onsite terms, the pairing terms including NN and LR interaction, and the inter-leg coupling terms, mathematically written as follows: 

\begin{equation}
\begin{aligned} 
H = & - \sum_{i=1}^{L-1} \Big[ t \big( c_{i}^{\dagger} c_{i+1} + d_{i}^{\dagger} d_{i+1} + \text{H.c.} \big) - 2\mu \sum_{i=1}^{L} \big( c_{i}^{\dagger} c_{i} + d_{i}^{\dagger} d_{i} \big)  \\ &+ \sum_{i=1}^{L-1} \Big[ \Delta \big( c_{i} c_{i+1}  + \sum_{i<j}^{L-j} \Big[ \frac{1}{\delta_l^{\alpha}} \big( d_{i} d_{j} + \text{H.c.} \big)  \Big] \big)  \Big] \\ & \; + t_v  \sum_{i=1}^{L}\big( c_{i}^{\dagger} d_{i} + \text{H.c.} \big) \Big]. 
\end{aligned}  
\label{H_vertical_LR_b_chain}
\end{equation}

Here, $t_\nu$ denotes the inter-leg coupling strength that simultaneously governs hopping between legs of the ladder.

\begin{figure}
\includegraphics[width=0.49\linewidth,height=0.5\linewidth]{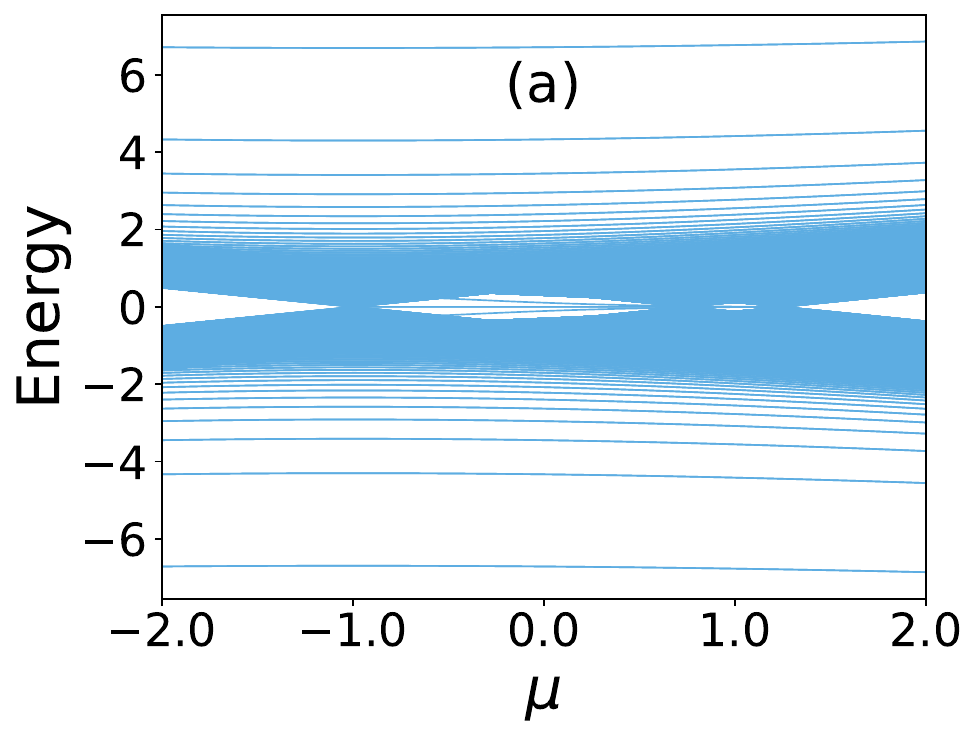}
\includegraphics[width=0.49\linewidth,height=0.5\linewidth]{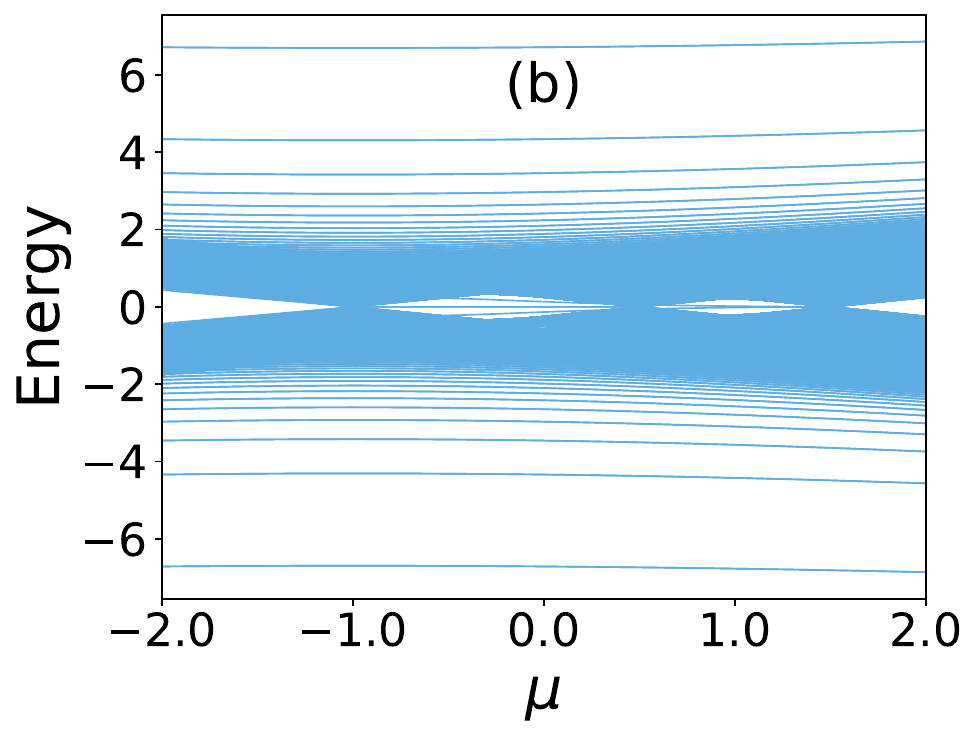}\\
\includegraphics[width=.7\linewidth,height=0.5\linewidth]{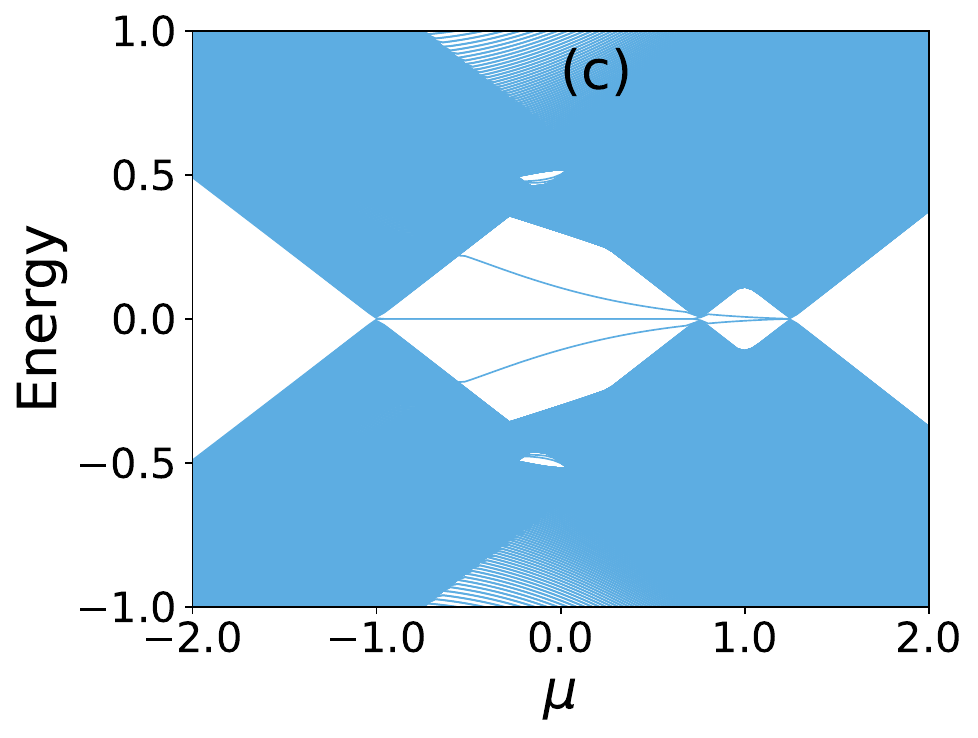}
\caption{Energy spectrum for the hybrid Kitaev ladder under OBC with $t=\Delta=1.0$, $\alpha=0.5$, and chain lengths $N=2L=1000$. The spectra are shown for different inter-leg coupling strengths: (a) $t_{\nu}=0.5$ and (b) $t_{\nu}=1.0$ (c) Zoomed-in view of the region of interest of energy spectrum shown in (a).}
\label{EG_HKC}
\end{figure}

\begin{figure}
\includegraphics[width=0.98\linewidth,height=0.5\linewidth]{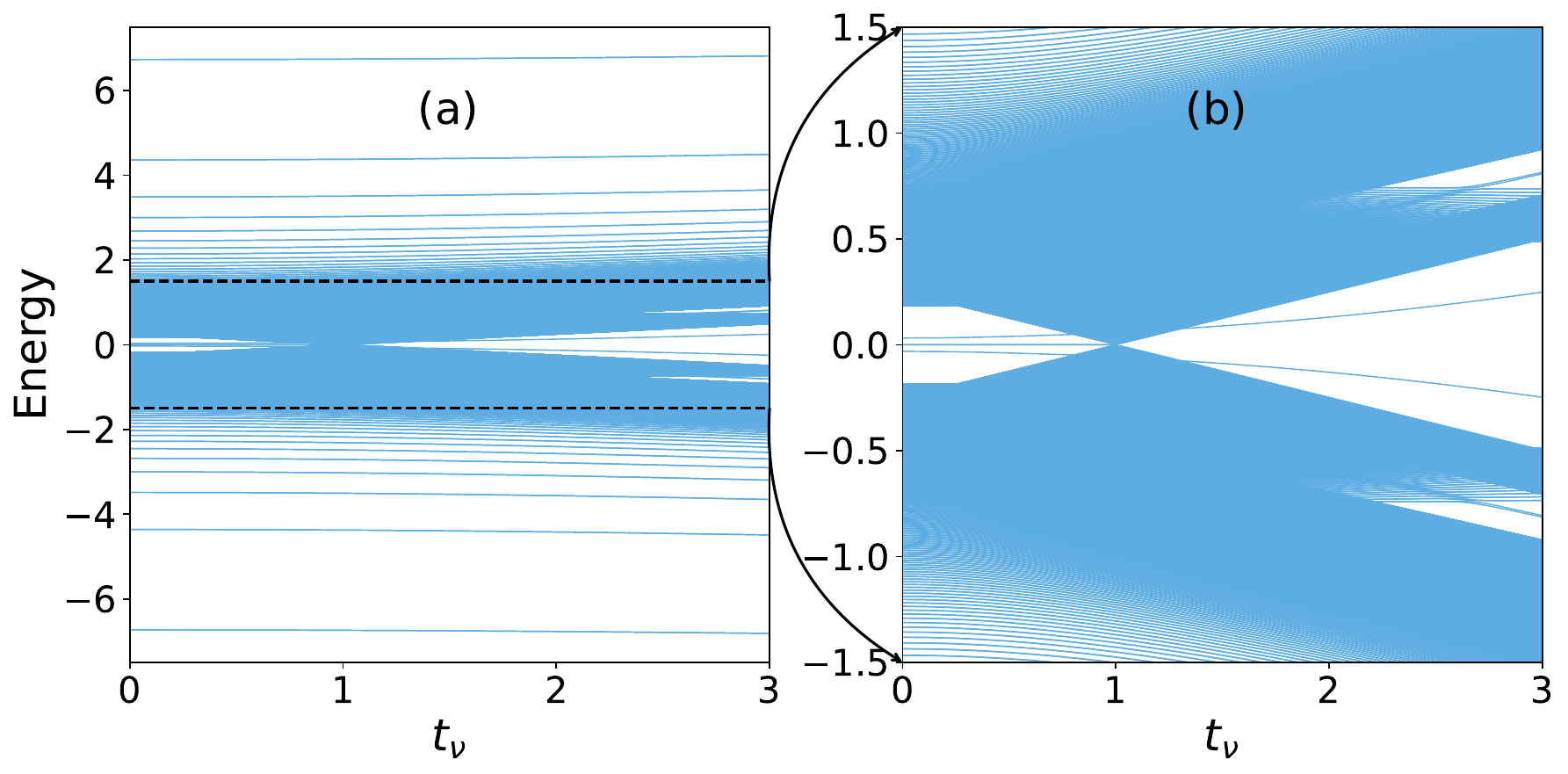}
\caption{(a) Energy spectrum as a function of the inter-leg coupling strength $t_{\nu}$ for the hybrid Kitaev ladder under OBC with fixed $\mu=0.5$, $t=\Delta=1.0$, and LR pairing exponent $\alpha=0.5$. (b) Zoomed in portion of (a) is shown for clarity.}
\label{EG_ES_HKC_tnu}
\end{figure}

\subsubsection*{\textbf{1. Energy Spectrum for the hybrid Kitaev ladder}}
\label{ES1}

First of all, we start with the suppressed inter-leg coupling strength, \( t_{\nu} = 0 \), where the hybrid Kitaev ladder is simply two independent chains: the Kitaev chain and an LR Kitaev chain. The individual energy spectrum of the Hamiltonian is already shown in Fig.~\ref{EG_KC_LKC}(a) for the conventional Kitaev chain and in Fig.~\ref{EG_KC_LKC}(b) for the LR Kitaev chain. The combined energy spectrum [Fig.~\ref{EG_KC_LKC}(c)] displays a topological phase in the interval \(-t < \mu < t\), with the presence of edge modes such as MZMs that coexist with finite energy MDMs. Outside of this region \(\mu < -t\) and \(\mu > t\), the system reveals in a trivial phase that does not support edge modes. Now, we introduce the inter-leg coupling strength: $t_{\nu} = 0.5$ as weak coupling strength and $t_\nu=1.0$ as normal coupling strength and the corresponding energy spectra are shown in Fig.~\ref{EG_HKC}(a) and (b), respectively.
\par
Upon turning on the inter-leg coupling $t_{\nu}$,  the topological regions associated with the edge modes are significantly modified. In the weak coupling regime $t_{\nu}=0.5$, the conventional topological transition point at $\mu=t$, present in both the Kitaev chain and the LR Kitaev chain, disappears and splits into two new transition points located approximately at \(\mu \simeq t-t_\nu/2\) and \(\mu \simeq t+t_\nu/2\) corresponding to $\mu \simeq 0.75$ and $\mu \simeq 1.25$, respectively.  In particular, the transition at \(\mu \simeq t+t_\nu/2\) marks the onset of a fully gapped phase dominated by MDMs. The interval \(-t < \mu < t-t_\nu/2\) supports the coexistence of MZMs and MDMs and \(-t-t_\nu/2 < \mu < t+t_\nu/2\) supports the MDMs, [see Fig.~\ref{EG_HKC}(c) which shows the zoomed in portion of Fig.~\ref{EG_HKC}(a)] indicating an asymmetric topological phase induced by inter-leg coupling. In contrast, the transition point at $\mu=-t$ remains nearly unchanged. The regions $\mu<-t$ and $\mu>t+t_\nu/2$ correspond to trivial phases with no edge modes. A qualitatively similar behaviour is observed for normal inter-leg coupling \(t_v = 1.0\) as shown in Fig.~\ref{EG_HKC}(b) the transition point $\mu = t$ is further shifted according to \(\mu \simeq t-t_\nu/2\) and \(\mu \simeq t+t_\nu/2\), leading to a reduction of the topological region supporting MZMs and MDMs. The interval  \(t-t_\nu/2 < \mu < t+t_\nu/2\) supports only MDMs, whereas the MZMs gradually disappear. These result demonstrate that increasing $t_{\nu}$ progressively suppresses and asymmetrically reshapes the topological phase, consistent with enhanced MBSs hybridization.
\par
Fig.~\ref{EG_ES_HKC_tnu} presents the energy spectrum with the inter-leg coupling strength $t_{\nu}$ for a particular chemical potential $\mu = 0.5$. We analyze the presence of MBSs, as MZMs and MDMs with $t_{\nu}$ continuously varied. The MBSs of the Kitaev chain, i.e., the MZMs persist up to $t_{\nu} \approx 1.0$, beyond which they disappear, indicating a transition. The MBSs of the LR Kitaev chain, i.e., the MDMs are clearly resolved for $t_{\nu} \lesssim 0.9$, after which they merge into the bulk spectrum. Interestingly, for $t_{\nu} \gtrsim 1.1$, MDMs re-emerge in the spectrum. These MBSs, the zero and finite energy bulk excitations of the energy spectrum for the hybrid Kitaev ladder Fig.~\ref{EG_HKC} and Fig.~\ref{EG_ES_HKC_tnu}  are confirmed to each other for the particular inter-leg coupling strength $t_{\nu}$.

\begin{figure}
\includegraphics[width=0.49\linewidth,height=0.5\linewidth]{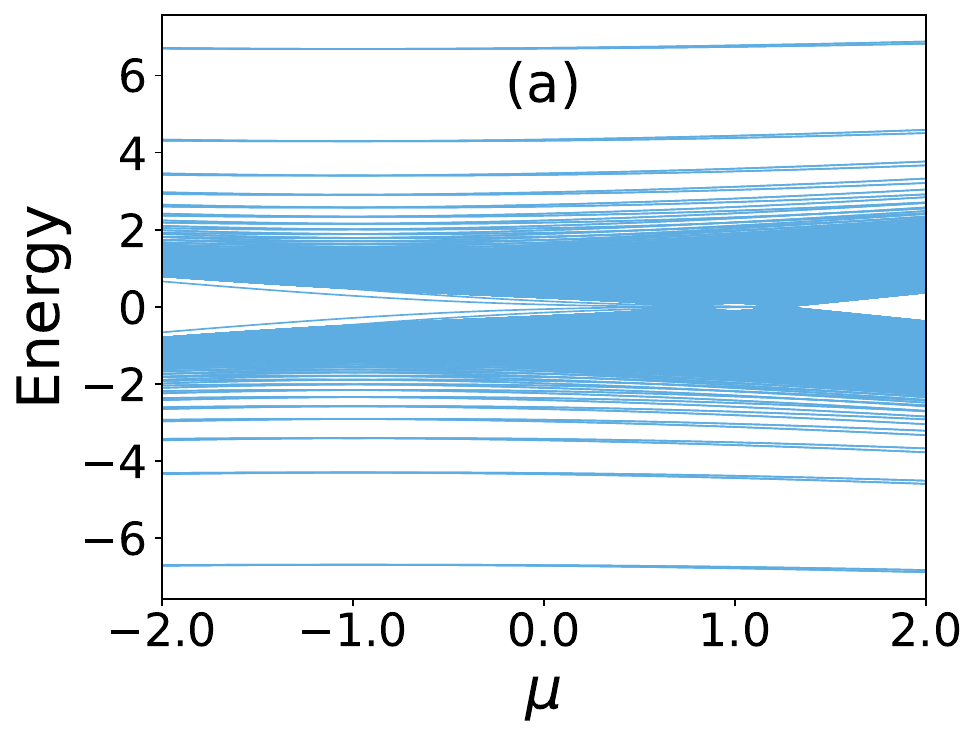}
\includegraphics[width=0.49\linewidth,height=0.5\linewidth]{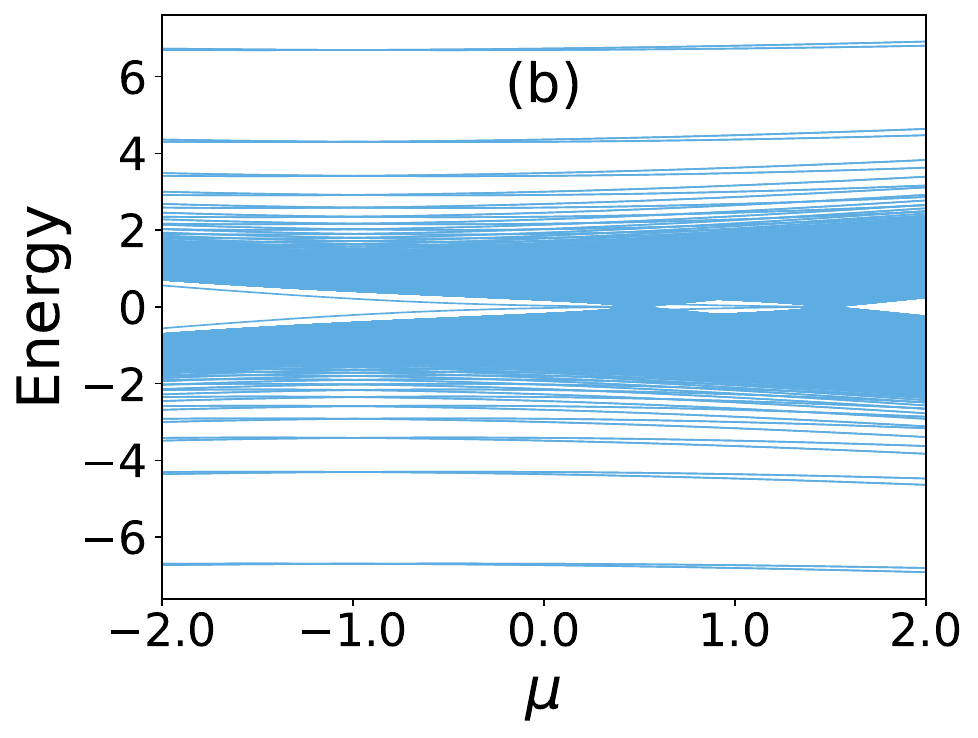}
\includegraphics[width=.7\linewidth,height=0.5\linewidth]{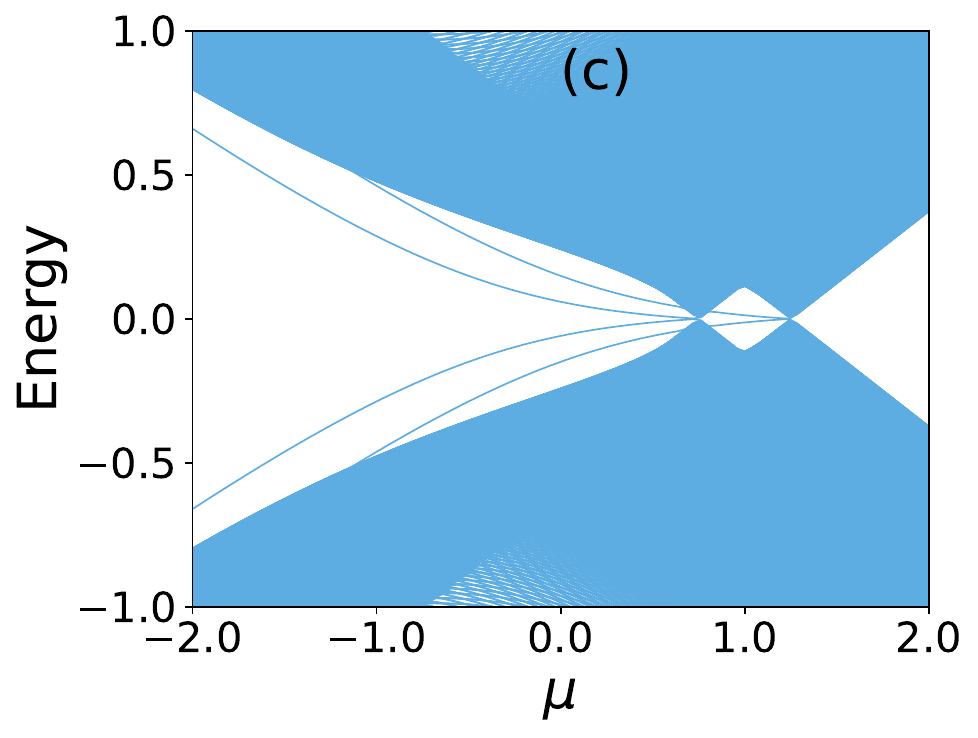}
\caption{Energy spectrum for the LR--LR Kitaev ladder under OBC with $t=\Delta=0.5$, $\alpha=0.5$, and chain lengths $N=2L=1000$. The spectra are shown for different inter-leg coupling strengths: (a) $t_{\nu}=0.5$ and (b) $t_{\nu}=1.0$ (c) Zoomed-in view of the region of interest of energy spectrum shown in (a).}
\label{EG_HKC1}
\end{figure}
\subsubsection*{\textbf{2. Energy Spectrum of LR-LR Kitaev ladder and NN-NN Kitaev ladder}}
\label{ES2}

We now discuss a more general realization of the hybrid Kitaev ladder [Eq.~(\ref{H_vertical_LR_b_chain})], namely the LR-LR Kitaev ladder, where both legs host the LR pairing with the same exponents $\alpha$. We also consider its limiting case corresponding to the NN-NN Kitaev ladder for $\alpha>1$. We start by analyzing the LR-LR Kitaev ladder, in which the topological regions are significantly modified compared to both the isolated LR Kitaev chain and the hybrid Kitaev ladder.

Fig.~\ref{EG_HKC1}(a) shows the energy spectrum for weak inter-leg coupling strength $t_{\nu}=0.5$. In this regime, the conventional topological transition point $\mu=t$ of the LR Kitaev chain disappears, and the inter-leg coupling generates new transition points located approximately at $\mu \simeq t \pm \frac{t_{\nu}}{2}$. These transition points separate distinct topological regions characterized by different numbers of MDMs. In particular, the interval $\mu < t-\frac{t_{\nu}}{2}$ supports four MDMs, while the region $t-\frac{t_{\nu}}{2} < \mu < t+\frac{t_{\nu}}{2}$ supports two MDMs [see Fig.~\ref{EG_HKC1}(c)]. Qualitatively similar behaviour is observed for normal inter-leg coupling $t_{\nu}=1.0$, shown in Fig.~\ref{EG_HKC1}(b). In this case, the shifted transition points at $\mu \simeq t \pm \frac{t_{\nu}}{2}$ again separate regions hosting four and two MDMs, respectively, with modified topological boundaries. The LR-LR Kitaev ladder therefore exhibits spectral features similar to those of the LR Kitaev chain, where the topological transition occurs predominantly for positive values of the chemical potential.
\par

\begin{figure}[!t]
\includegraphics[width=0.49\linewidth,height=0.5\linewidth]{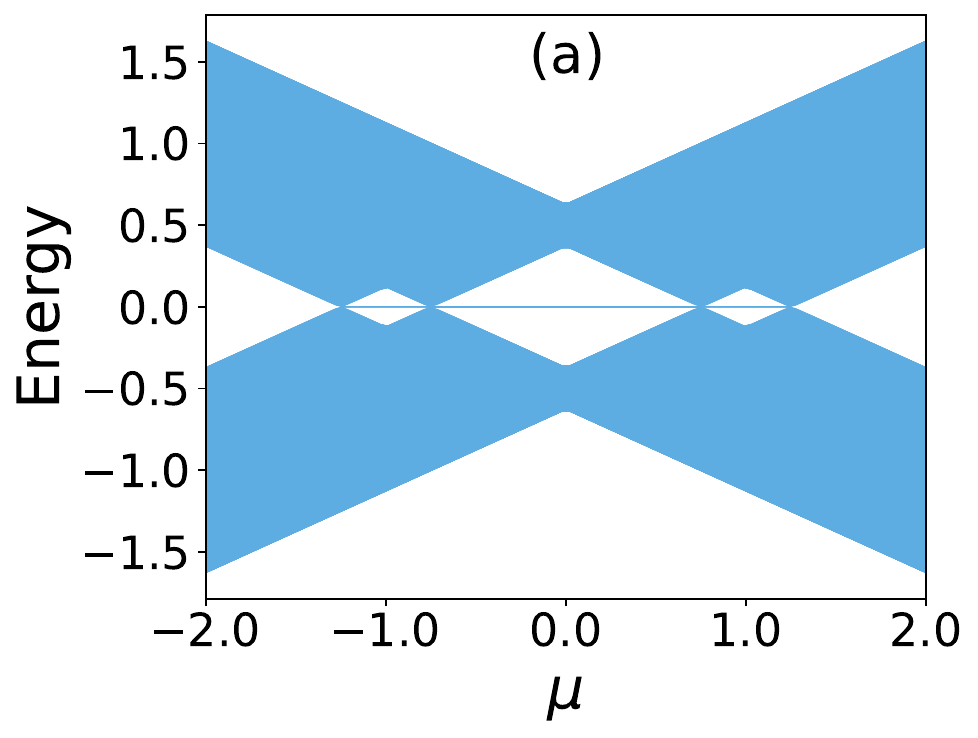}
\includegraphics[width=0.49\linewidth,height=0.5\linewidth]{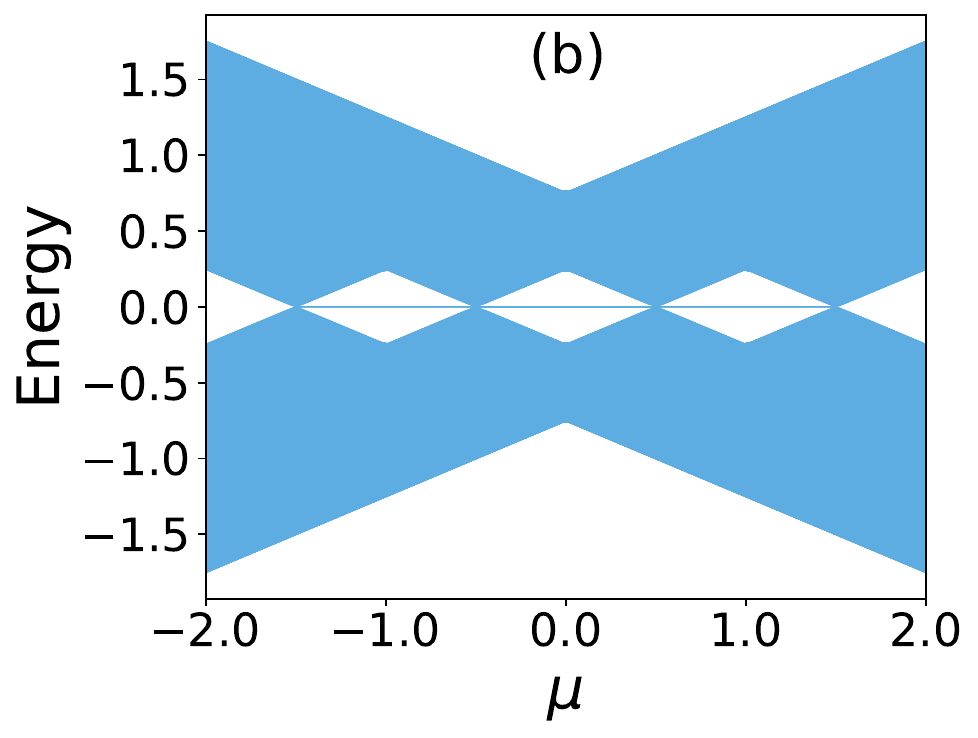}
\caption{Energy spectrum for the NN - NN Kitaev ladder under OBC with $t=\Delta=1.0$ and chain lengths $N=2L=1000$. The spectra are shown for different inter-leg coupling strengths: (a) $t_{\nu}=0.5$ and (b) $t_{\nu}=1.0$.}
\label{EG_HKC2}
\end{figure}

We now discuss the NN-NN Kitaev ladder, where both legs host normal pairing. Fig.~\ref{EG_HKC2}(a) shows the energy spectrum for weak inter-leg coupling strength $t_{\nu}=0.5$. In this regime, the conventional topological transition points at $\mu=\pm t$, characteristic of the Kitaev chain, disappear due to inter-leg coupling $t_{\nu}$. The $t_{\nu}$ generates new transition points approximately located at $\mu \simeq \pm t \pm \frac{t_{\nu}}{2}$. These points separate distinct topological regions characterized by different numbers of MZMs. In particular, the central region $-t+\frac{t_{\nu}}{2} < \mu < t-\frac{t_{\nu}}{2}$ supports four MZMs, while the outer regions $- t-\frac{t_{\nu}}{2} < \mu < -t+\frac{t_{\nu}}{2}$ and $t-\frac{t_{\nu}}{2} < \mu <  t+\frac{t_{\nu}}{2}$ support two MZMs. For $\mu>|t+t_\nu/2|$ the system is revealed in a trivial phase. Qualitatively similar behaviour is observed for normal inter-leg coupling $t_{\nu}=1.0$, shown in Fig.~\ref{EG_HKC2}(b). In this case, the shifted transition points at $\mu \simeq \pm t \pm \frac{t_{\nu}}{2}$ again separate regions hosting four and two MZMs, respectively, with modified topological boundaries.  Beyond $|\mu| > t+\frac{t_{\nu}}{2}$, the system reveals a trivial phase. Overall, the energy spectrum of the NN--NN Kitaev ladder exhibits qualitative features analogous to those of the conventional Kitaev chain, with topological transitions with all values of the chemical potential.
\par
The results presented here demonstrate that the nature of the MBSs and the associated topological regions are strongly influenced by the exponent $\alpha$ in the hybrid Kitaev ladder. In the LR--LR Kitaev ladder, the LR pairing significantly modifies the topological phases characterized primarily by MDMs  with respect to the chemical potential. In contrast, the NN--NN Kitaev ladder retains the conventional physics of the Kitaev model, exhibiting topological phases with four and two MZMs with respect to the chemical potential. The inter-leg coupling $t_{\nu}$ plays a crucial role in both systems by shifting the phase boundaries and modifying the stability regions of the edge modes. These observations highlight the distinct impact of $\alpha$ on the topological properties and establish a clear connection between the pairing range, the structure of the MBSs, and the evolution of the phase boundary.

\begin{figure}
\includegraphics[width=1.0\linewidth,height=0.8\linewidth]{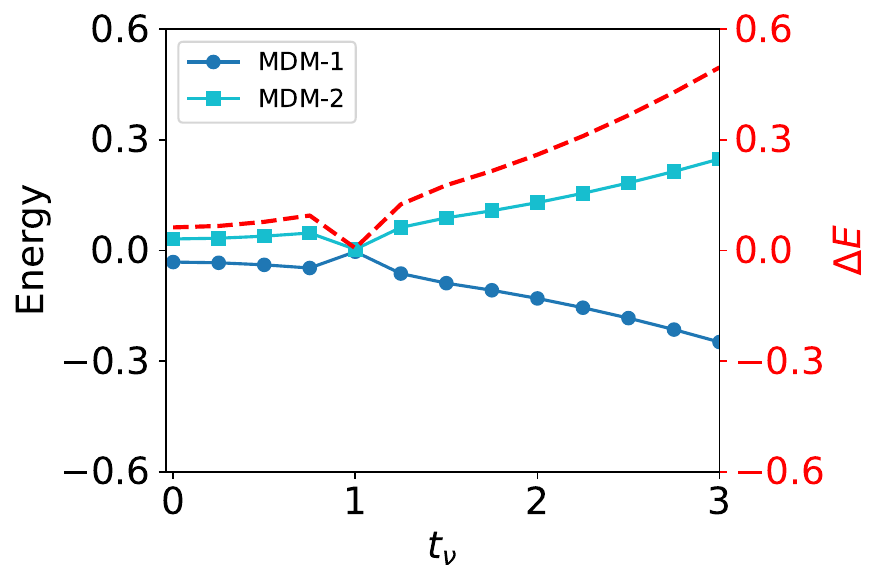}
\caption{Bulk excitation MDMs energy and their gap $\Delta E$ versus the inter-leg coupling strength $t_\nu$ for $L_1=500,L_2=500, \alpha=0.5$ with chemical potential $\mu = 0.5$.}
\label{EG_MZM}
\end{figure}

\subsection*{C. Gap evolution with inter-leg coupling for modes}

We now investigate the effect of the inter-leg coupling \(t_\nu\) on the MBSs of fixed chemical potential in a hybrid Kitaev ladder. We analyze the evolution of MDMs [Fig.~\ref{EG_MZM}], as a function of \(t_\nu\). These MBSs of the hybrid Kitaev ladder are shown in the spectrum [Fig.~\ref{EG_HKC} and Fig.~\ref{EG_ES_HKC_tnu}]. In addition, we define the energy gap for the MDMs as
\begin{equation}
\Delta E = E_{\mathrm{MDM1}} - E_{\mathrm{MDM2}},
\end{equation}
which provides a direct measure of the bulk excitation gap. As \(t_\nu\) increases, both MDMs shift monotonically in energy, reflecting the enhanced bulk hybridization induced by inter-leg coupling. As a result, the excitation gap \(\Delta E\) increases with \(t_\nu\), indicating a progressive opening of the MDMs gap. This behaviour demonstrates that \(t_\nu\) primarily renormalizes the bulk sector of the spectrum while preserving the overall gap nature of the system. In particular, the MZMs remain pinned at zero energy and survive up to at least $t_{\nu}=1.0$ for $\mu=0.5$, demonstrating the robustness of the topological edge states against moderate inter-leg hybridization.
\par
The energy spectra discussed above reveal a rich topological structure governed by the interplay of $t_\nu$ and $\alpha$. In addition, we now turn to the momentum-space formulation of the hybrid Kitaev ladder. By analyzing the symmetry properties and bulk dispersion of the Bogoliubov-de Gennes (BdG) Hamiltonian, we identify the symmetry class and  the gap-closing conditions for the observed topological phase transitions.

\begin{figure}
\includegraphics[width=0.48\linewidth,height=0.4\linewidth]{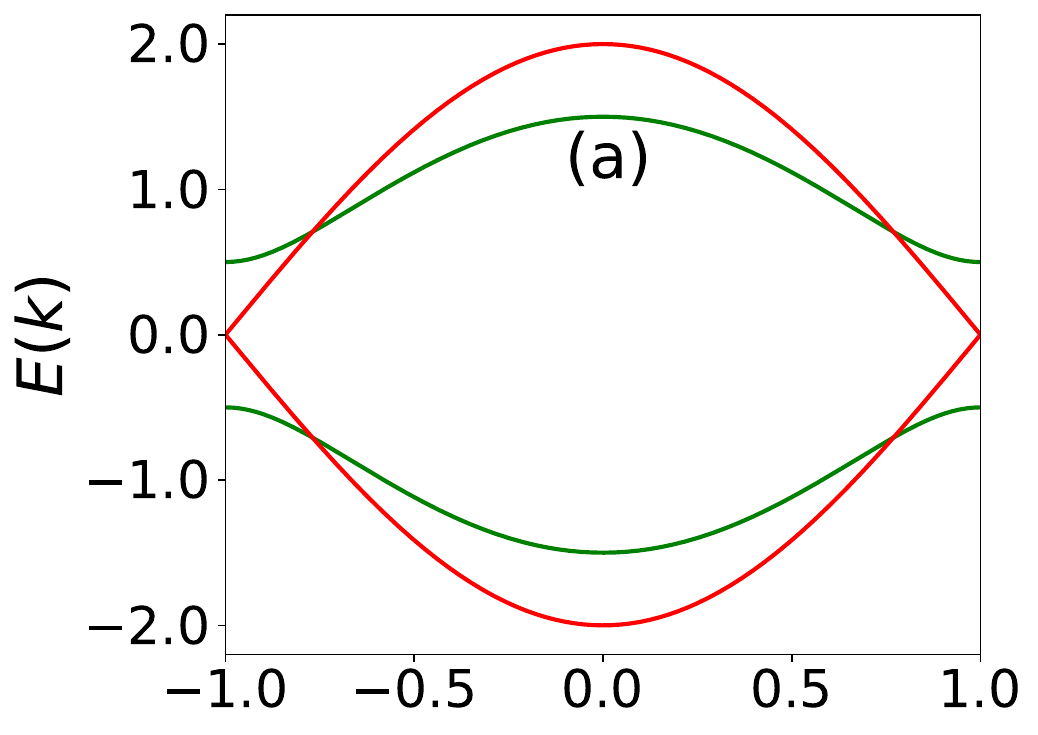}
\includegraphics[width=0.49\linewidth,height=0.4\linewidth]{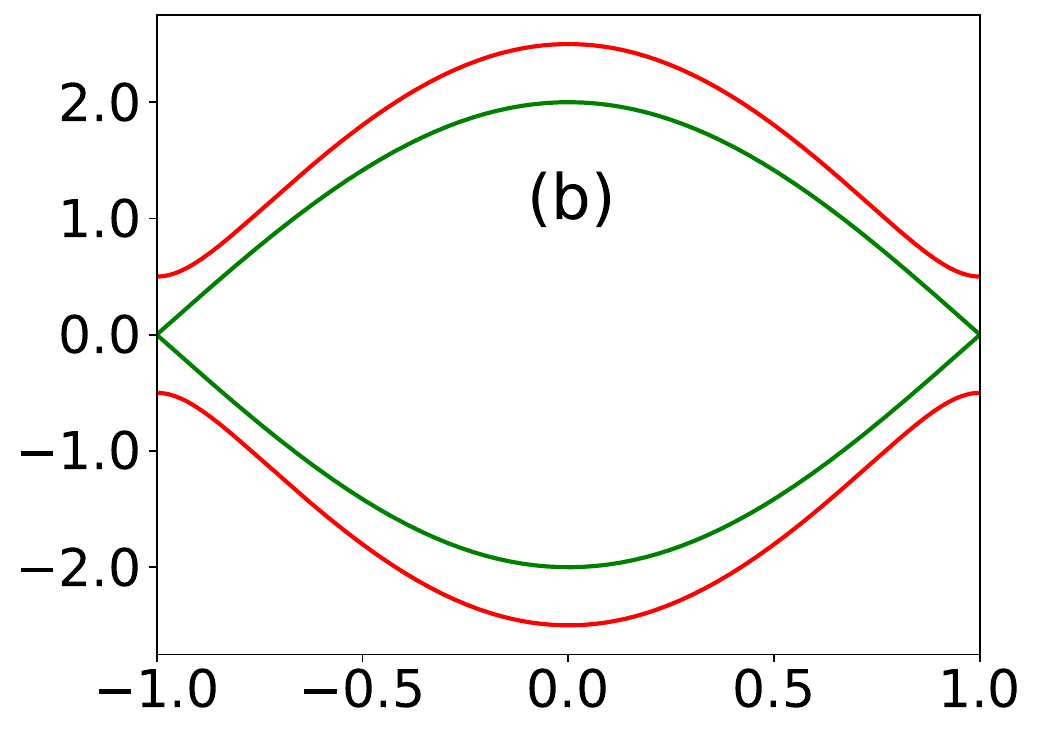}
\includegraphics[width=0.49\linewidth,height=0.4\linewidth]{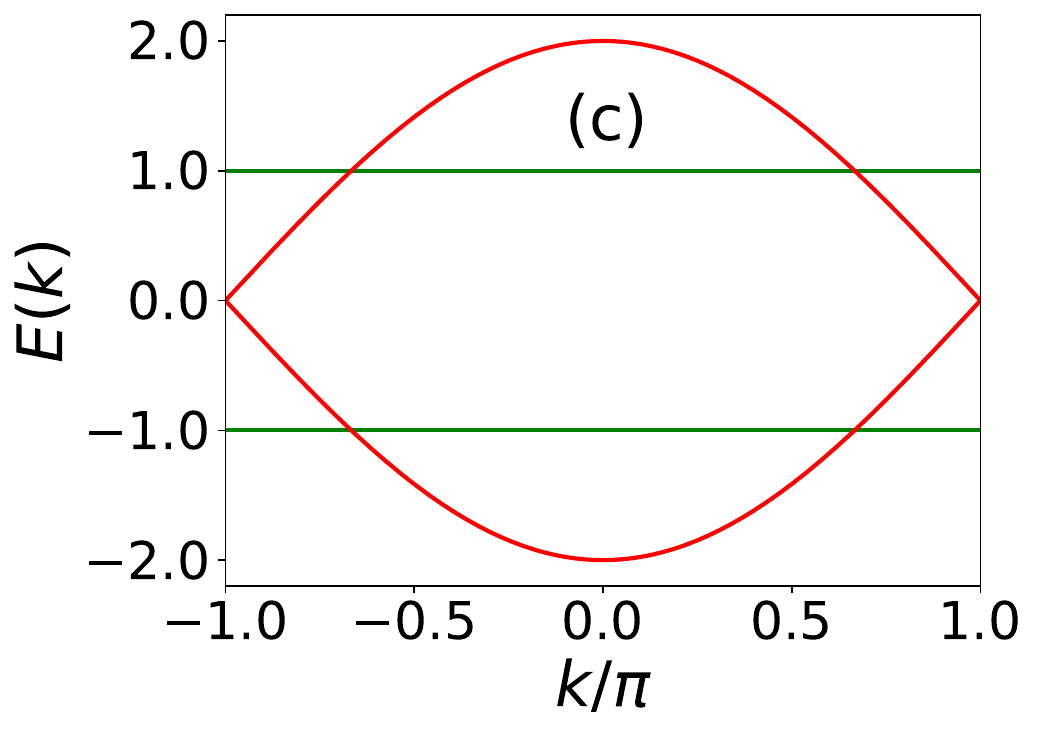}
\includegraphics[width=0.49\linewidth,height=0.4\linewidth]{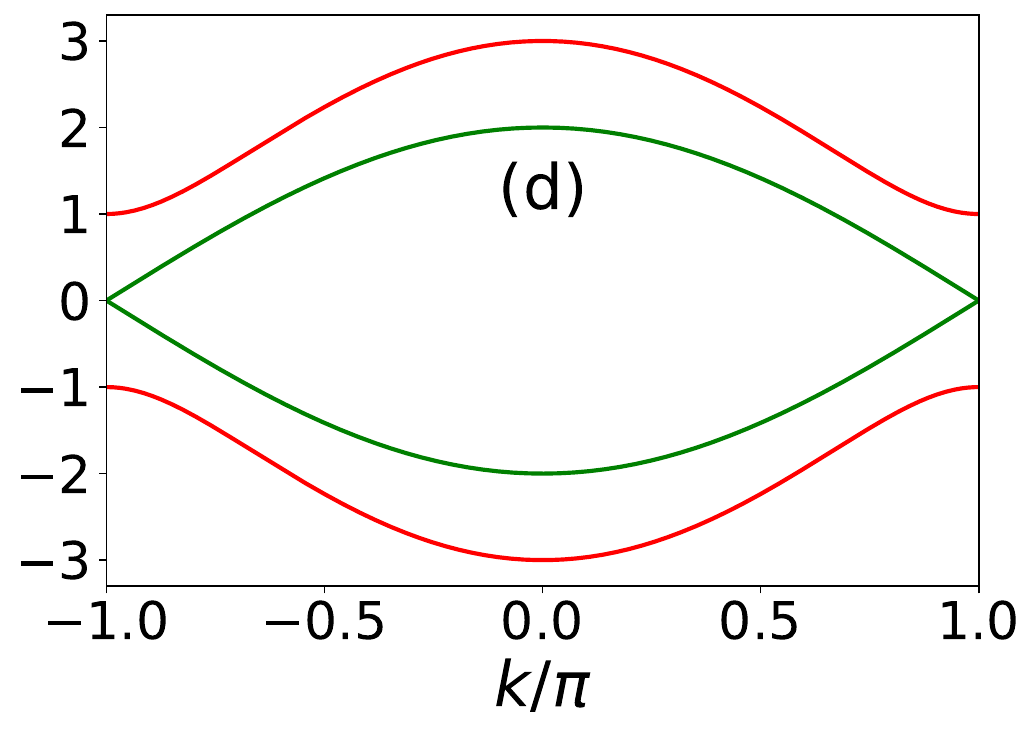}
\caption{Dispersion relations for the NN-NN Kitaev ladder with $t = \Delta = 1.0$. (a,b) $t_{\nu} = 0.5$ with $\mu$ (a) $0.75$ and (b) $1.25$; (c,d) $t_{\nu} = 1.0$ with $\mu$ (c) $0.5$ and (d) $1.5$. Gap closings in the spectrum indicate topological phase transitions.}
\label{EG_Dispersion}
\end{figure}

%%%%%%%%%%%%%%%%%%%%%%%%%%%%%%%%%%%%%%%%%%%%%%%%%%%%%%%%%%%%
\section{Symmetry and Dispersion relation}
\label{Symmetry_DR}

By imposing periodic boundary conditions, the hybrid Kitaev ladder Hamiltonian [Eq.~(\ref{H_vertical_LR_b_chain})] can be transformed into momentum space through the Fourier transformation. In the Nambu basis \( \left(c_{k}^\dagger,\, d_{k}^\dagger,\, c_{-k},\, d_{-k}\right)^T, \) the BdG Hamiltonian can be written as
\begin{equation}
H(k) =
\begin{pmatrix}
\epsilon_k & t_{\nu}/2 & \Delta_{Nk} & 0 \\
t_{\nu}/2 & \epsilon_k & 0 & \Delta_{Lk} \\
\Delta_{Nk}^{*} & 0 & -\epsilon_k & -t_{\nu}/2 \\
0 & \Delta_{Lk}^{*} & -t_{\nu}/2 & -\epsilon_k
\end{pmatrix},
\label{BdG_k}
\end{equation}
where
\begin{equation}
\epsilon_k = - t\cos(k) - \mu, 
\quad
\Delta_{Nk} = i\Delta\, \sin{k},
\quad
\Delta_{Lk} = i\Delta\, f_\alpha(k),
\end{equation}
with
\begin{equation}
f_\alpha(k)=\sum_{l=1}^{\infty}\frac{\sin(kl)}{l^\alpha},
\end{equation}
characterizes the LR pairing between sites $l=|i-j|$ in momentum space.

The BdG Hamiltonian [Eq.~(\ref{BdG_k})] preserves the PH symmetry,
\begin{equation}
\sigma_x H^{T}(k)\sigma_x = -H(-k),
\end{equation}
with the anti-unitary particle-hole operator given by
\begin{equation}
\mathcal{C}=\sigma_x \mathcal{K},
\end{equation}
where $\mathcal{K}$ represents the complex-conjugation. The BdG Hamiltonian also preserves TR symmetry,
\begin{equation}
H^{*}(k)=H(-k),
\end{equation}
which is represented by the anti-unitary operator,
\begin{equation}
\mathcal{T}=\mathcal{K}.
\end{equation}
As a consequence of the coexistence of the PH and TR symmetry, the system additionally possesses chiral symmetry,
\begin{equation}
\sigma_x H(k)\sigma_x = -H(k).
\end{equation}
This implies that the hybrid Kitaev ladder belongs to the BDI symmetry class. In the limiting case \(\alpha>1\), the hybrid Kitaev ladder effectively becomes the NN-NN Kitaev ladder and the corresponding dispersion relation takes a simpler form as 
\begin{equation}
E(k) = \pm
(\epsilon_k \pm t_{\nu}/2)^2 + |\Delta_{Nk}|^2,
\label{dispersion_final_NN}
\end{equation}

The dispersion relation [Eq.~(\ref{dispersion_final_NN})] for the LR exponent \(\alpha>1\) is shown in Fig.~\ref{EG_Dispersion}. For weak inter-leg coupling strength $t_{\nu} = 0.5$, the dispersion relation in $\mu = t-t_\nu/2$ and $\mu = t+t_\nu/2$ is presented in Figs.~\ref{EG_Dispersion} (a,b), respectively. Although for normal inter-leg coupling strength $t_{\nu} = 1.0$, the dispersion relation exhibits the same qualitative features as those observed for $t_{\nu}=0.5$. The transition points of the bulk gap $\mu=t-t_{\nu}/2$ and $\mu=t+t_{\nu}/2$, are illustrated in Figs.~\ref{EG_Dispersion}(c,d), respectively. As $\mu$ is tuned, the energy gap closes at specific momenta, leading to gapless excitations. These gap closure points signal topological phase transition points of the hybrid Kitaev ladder for the LR exponent $\alpha>1.0$.
\par
We find that the transition points obtained from the energy spectrum in Fig.~\ref{EG_HKC2} coincide with the bulk-gap closing points identified in the dispersion relation shown in Fig.~\ref{EG_Dispersion}. This correspondence confirms that the topological phase transitions of the hybrid Kitaev ladder originate from the closing and reopening of the bulk excitation gap.
In the absence of inter-leg coupling ($t_{\nu}=0$), the  bulk gap closes at $k=0$, giving the well-known transition points $\mu=\pm t$ for the Kitaev chain. When $t_{\nu}$ is introduced for $\alpha>1$, the two legs hybridize and the transition point splits into $\mu \pm \frac{t_{\nu}}{2}$ equivalent to $\pm t \pm \frac{t_{\nu}}{2}$ and the corresponding dispersion are gapless. The splitting of the transition point into two distinct phase boundaries is in good agreement in the energy spectra [Figs.~\ref{EG_HKC2}] and phase diagrams shown later.
\par
The BDI symmetry in the topological phases for hybrid Kitaev ladder permits multiple MBSs, whose number depends sensitively on the chemical potential, inter-leg coupling strength, and LR exponent. Having established these analytical insights, we now turn to the phase diagram, where the evolution of the topological phases and their associated edge modes are systematically mapped in the relevant parameters in the hybrid Kitaev ladder.

\begin{figure}
\includegraphics[width=0.49\linewidth,height=0.48\linewidth]{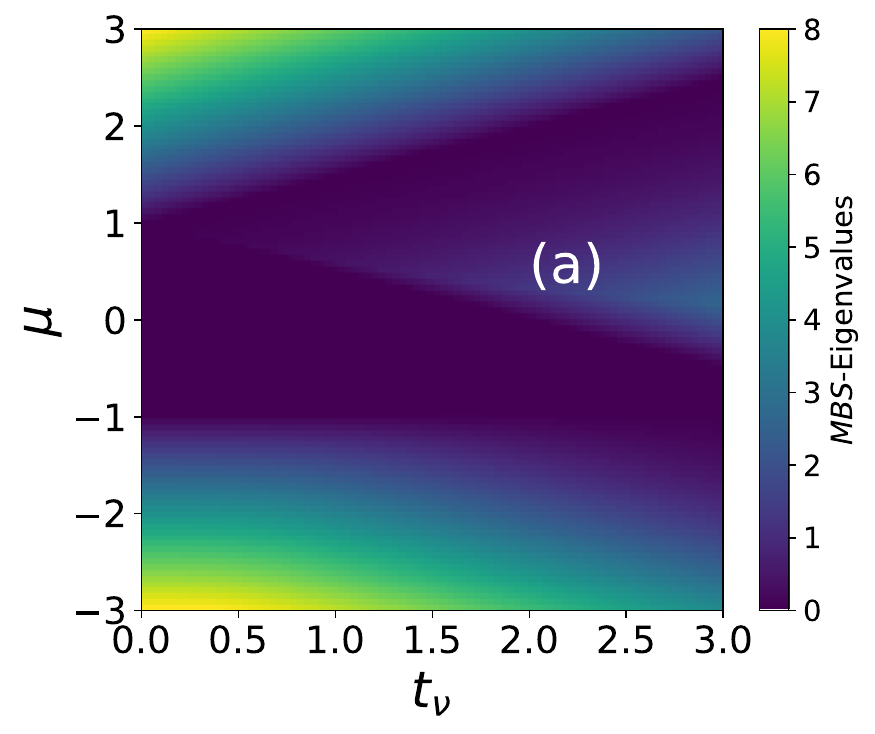}
\includegraphics[width=0.49\linewidth,height=0.48\linewidth]{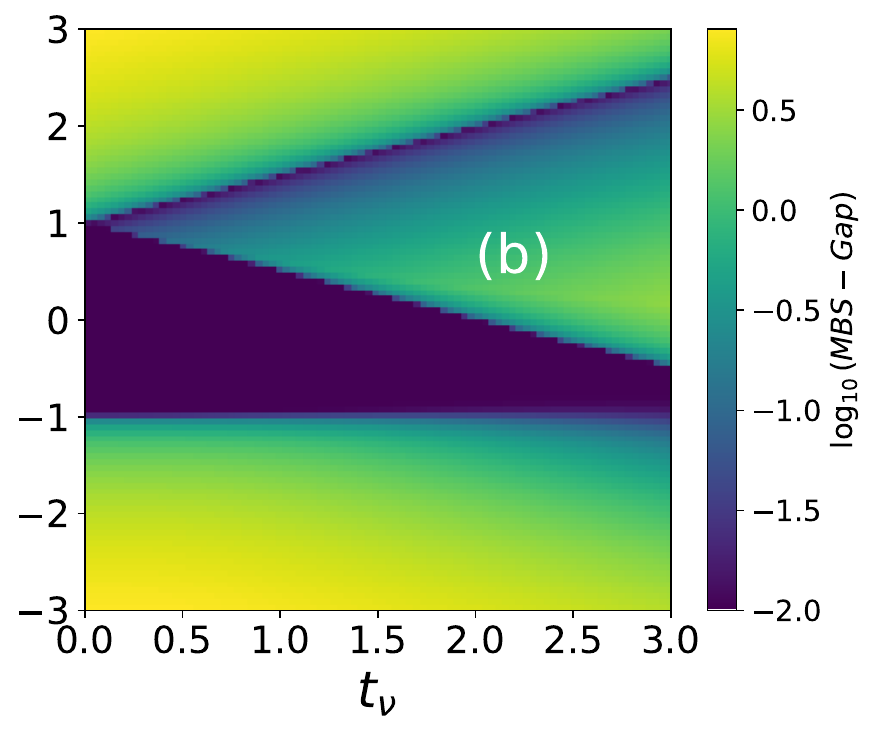}
\caption{Phase diagram of the hybrid Kitaev ladder showing the evolution of (a)  zero (MZMs) and finite energy (MDMs) modes (b)  corressponding bulk excitation gap $(\log_{10}(MBS - Gap))$, as a function of $\mu$ and $t_{\nu}$. The dark region corresponds to gap closing, indicating the presence of MZMs and MDMs, the bright region inside the topological phase indicates the presence of MDMs.  Parameters: $t=\Delta=1.0$ and $\alpha=0.5$.}
\label{EG_heat_phase}
\end{figure}

\section{Phase Diagram}
\label{PD_HKL}

The phase diagram of the hybrid Kitaev ladder in the $(\mu,t_{\nu})$ plane is shown in Fig.~\ref{EG_heat_phase}, illustrating the evolution of the topological phase boundary with the inter-leg coupling strength $t_{\nu}$. As $t_{\nu}$ increases, the transition point initially located at $\mu=t$ shifts monotonically toward higher and lower values of $\mu$, indicating that $t_{\nu}$ stabilizes the topological phase. In contrast, the other transition point at $\mu=-t$ remains nearly unchanged with increasing $t_{\nu}$, reflecting the asymmetric role of the LR pairing in the topological phase. In the absence of $t_{\nu}$, the topological phase of the NN Kitaev chain is determined by the condition $|\mu|<t$. However, in the hybrid Kitaev ladder, $t_{\nu}$ hybridizes the two chains and effectively splits the low-energy spectrum, leading to shifted transition points and modified topological boundaries. The evolution of the spectrum in Fig.~\ref{EG_heat_phase}(a) demonstrates that the topological transition is accompanied by the closing and reopening of the bulk excitation gap. The corresponding heat map of the excitation gap in the $(\mu,t_{\nu})$ plane, as shown in Fig.~\ref{EG_heat_phase}(b), clearly distinguishes the regions of the topological phases that host MBSs. The dark regions correspond to zero-energy modes associated with MZMs and finite-energy MDM, whereas the central brighter regions correspond to finite-energy MDM excitations. In Fig.~\ref{EG_heat_phase}(a,b), the progressive expansion and shift of the low-energy region with increasing $t_{\nu}$ are consistent with the evolution of the phase boundary observed in the spectrum.

\begin{figure}[t]
\centering
\begin{tikzpicture}[x=0.8cm, y=1.4cm]

\draw[->, thick] (-4.0,-1.0) -- (4.5,-1.0) node[right] {$\mu$};
\draw[->, thick] (0,0.0) -- (0,2.5) node[above] {$\alpha$};

\fill[black!80!black!70] (-3,0.5) rectangle (3,2.0);

\node at (0,1.7) {\textcolor{white}{(a)}};
\node at (0,1.2) {\textcolor{white}{Four MZM}};
\node at (2.0,1.2) {\textcolor{white}{Two MZM}};
\node at (-2.0,1.2) {\textcolor{white}{Two MZM}};

\fill[olive!80!black] (-3,0.3) rectangle (3,0.6);
\node at (0,0.45) {\textcolor{white}{Crossover Phase}};

\fill[red!80] (-3.0,-1.0) rectangle (3,0.3);

\node at (0,-0.3) {\textcolor{white}{MDM}};
\node at (0,-0.6) {\textcolor{white}{MZM}};
\node at (-1.5,-0.3) {\textcolor{white}{MZM}};
\node at (2.0,-0.3) {\textcolor{white}{MDM}};

\fill[gray!60] (3,-1) rectangle (4.3,2.0);
\fill[gray!60] (-3.1,-1) rectangle (-2,0.32);
\fill[gray!60] (-3,-1) rectangle (-4,2);

\node at (3.7,1.2) {Trivial};
\node at (-3.25,0.0) {Trivial};

\draw[dashed, thick] (-2,-1.0) -- (-2,0.3);
\draw[dashed, thick] (1,-1) -- (1,0.3);
\draw[dashed, thick] (-1,-1) -- (-1,0.3);
\draw[dashed, thick] (1,0.4) -- (1,2.0);
\draw[dashed, thick] (-1,0.4) -- (-1,2.0);
\draw[dashed, thick] (3,-1) -- (3,2);
\draw[dashed, thick] (-3,0.35) -- (-3,2);

\node at (-3,-1.2) {$-1.5$};
\node at (-2,-1.2) {$-1.0$};
\node at (-1,-1.2) {$-0.5$};
%\node at (0,-1.2) {$0.0$};
\node at (1,-1.2) {$0.5$};
\node at (2,-1.2) {$1.0$};
\node at (3,-1.2) {$1.5$};

\node at (3.7,-0.3) {$\alpha = 0.5$};
\node at (3.7,0.45) {$\alpha = 1$};
\node at (3.7,1.5) {$\alpha = 1.5$};

\end{tikzpicture}
\includegraphics[width=0.85\linewidth,height=0.7\linewidth]{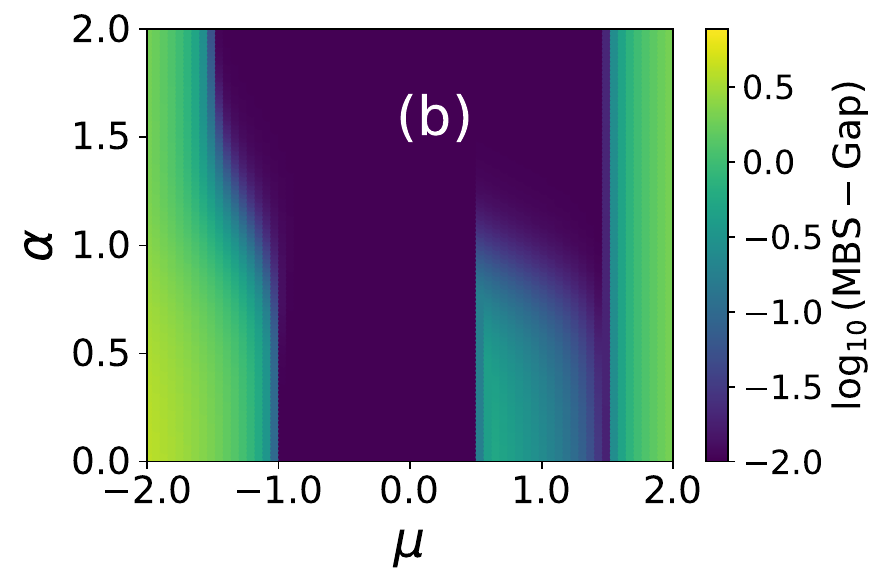}
\caption{(a) Schematic phase diagram of the hybrid Kitaev ladder in the $(\alpha,\mu)$ plane for fixed inter-leg coupling $t_{\nu}=1.0$. The system exhibits distinct topological phases characterized by different numbers of MZMs and MDMs. For $\alpha<1$, a two-MZM phase is realized, while for $\alpha>1$, a four-MZM phase emerges in topological phase. The intermediate region corresponds to a crossover regime where edge and bulk excitations coexist. For large $|\mu|$, the system becomes topologically trivial. Dashed lines indicate phase boundaries. (b) Numerically calculated phase diagram of the hybrid Kitaev ladder in the $(\alpha,\mu)$ plane for fixed inter-leg coupling $t_{\nu}=1.0$. The color scale represents $\log_{10}(MBS - Gap)$, where $\Delta E$ is the bulk excitation gap obtained from the BdG spectrum. Dark regions correspond to gap closing, indicating topological phase boundaries. The central gapped region represents to the topological phase hosting MZMs, while the outer regions represent trivial phases.}
\label{EG_HKC_PD}
\end{figure}
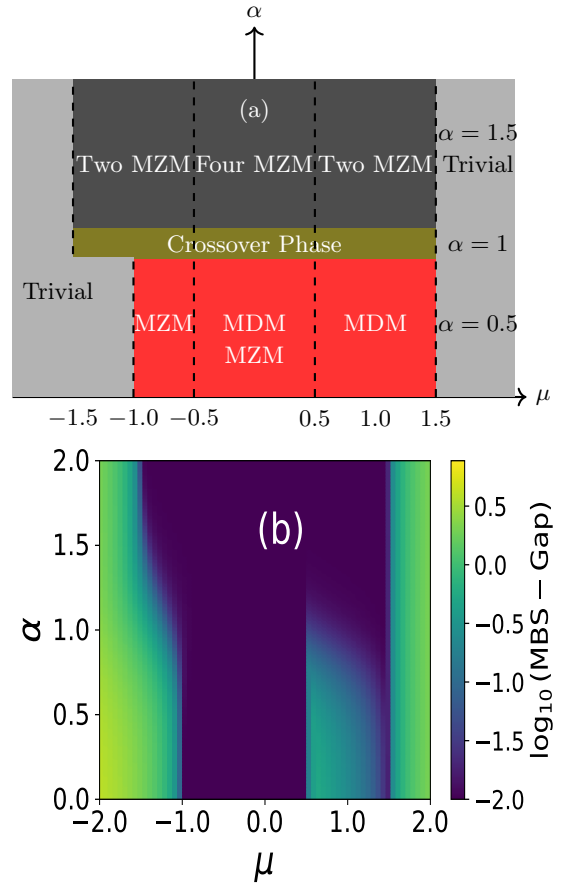

Fig.~\ref{EG_HKC_PD} illustrates the topological phase diagram of the hybrid Kitaev ladder in the $(\alpha,\mu)$ plane, illustrating the role of the LR pairing exponent $\alpha$ in the topological properties. The schematic phase diagram [Fig.~\ref{EG_HKC_PD}a] inferred from the energy spectrum [Figs. \ref{EG_HKC} and \ref{EG_HKC2}] reveals a rich topological property. A numerical phase diagram [Fig. \ref{EG_HKC_PD}(b)] is also obtained from the bulk excitation gap with the MBSs, confirming the identified topological regions. For $\alpha<1$, both phase diagrams exhibit asymmetric topological regions with respect to  $\mu \leftrightarrow -\mu$.  In this regime, the interval $-t-t_\nu/2<\mu<-t+t_\nu/2$ supports two MZMs, while the region $-t+t_\nu/2<\mu<t-t_\nu/2$ supports two MZMs that coexist with two MDMs. The interval $t-t_\nu/2<\mu<t+t_\nu/2$ supports only MDMs. In general, the topological region extends approximately over $-t < \mu < t+\frac{t_{\nu}}{2}$, demonstrating a pronounced asymmetry in the chemical potential induced by the LR pairing and inter-leg hybridization. As $\alpha$ increases, the system undergoes a crossover from the asymmetric phase to a symmetric topological regime about $\mu$. This transition is mediated by a coexistence region where signatures of both MZMs and MDMs are simultaneously present, reflecting enhanced hybridization between edge and bulk excitations. For $\alpha>1$, the system enters a higher topological phase characterized by four MZMs. In this regime, two finite-energy MDMs collapse into zero-energy states, resulting in an additional pair of MZMs. The topological region becomes symmetric under $\mu \leftrightarrow -\mu$ and extends over $-t-t_\nu/2<\mu<t+t_\nu/2$ Within this phase, the central region  $-t+t_\nu/2<\mu<t-t_\nu/2$ supports four MZMs, while the outer regions  $-t-t_\nu/2<\mu<-t+t_\nu/2$ and $t-t_\nu/2<\mu<t+t_\nu/2$ support two MZMs. For sufficiently large values of $|\mu|$, the system becomes topologically trivial, with no MBSs and a possessing fully gapped spectrum. 
\par
Interestingly, the numerical phase diagram shown [Fig.~\ref{EG_HKC_PD}(b)] is in excellent agreement with the schematic phase diagram [Fig.~\ref{EG_HKC_PD}(a)]. For $\alpha<1$, the numerical results clearly reproduce the predicted schematic asymmetric topological structure, where the system hosts two MZMs together with regions containing coexisting MZMs and MDMs. The dark regions in the numerically calculated gap map correspond to these low-energy topological sectors, while the outer bright regions represent fully gapped trivial phases. In this regime, the crossover region, where MZMs coexist with MDMs, is consistently observed in both diagrams. As the LR exponent increases beyond $\alpha>1$, the numerical phase diagram confirms the transition into a symmetric topological phase. In particular, the enlargement of the central topological region and the emergence of additional MZMs in the numerical spectrum are fully consistent with the schematic prediction. The numerically computed gap-closing lines accurately reproduce the phase boundaries separating the trivial, two-MZM, and four-MZM regions. This agreement between the schematic and numerical phase diagrams demonstrates the robustness of the topological phases and establishes the LR pairing exponent $\alpha$ as an effective tuning parameter for engineering distinct Majorana sectors in the hybrid Kitaev ladder.

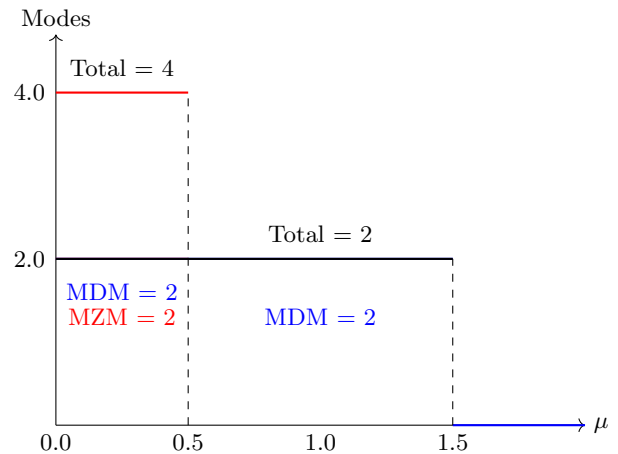
\begin{figure}[t]
\centering
\begin{tikzpicture}[x=3.5cm, y=1.1cm]

\draw[->] (0,0) -- (2.0,0) node[right] {$\mu$};
\draw[->] (0,0) -- (0,4.7) node[above] {Modes};

\draw[thick, red] (0,2) -- (0.5,2);
\draw[thick, red] (0,4) -- (0.5,4);

\node[red] at (0.25,1.3) {MZM = 2};
\node[blue] at (0.25,1.6) {MDM = 2};

\draw[thick, blue] (0,2) -- (1.5,2);
\draw[thick, blue] (1.5,0) -- (2.0,0);

\node[blue] at (1.0,1.3) {MDM = 2};
\node[black] at (1.0,2.3) {Total = 2};

\draw[thick, black] (0,2) -- (1.5,2);

\node at (0.25,4.3) {Total = 4};

\draw[dashed] (0.5,0) -- (0.5,4);
\draw[dashed] (1.5,0) -- (1.5,2);

\node at (0.0,-0.2) {$0.0$};
\node at (0.5,-0.2) {$0.5$};
\node at (1.0,-0.2) {$1.0$};
\node at (1.5,-0.2) {$1.5$};
\node at (-0.1,2.0) {$2.0$};
\node at (-0.1,4.0) {$4.0$};

\end{tikzpicture}
\caption{Schematic shows the variation of the number of MZMs, MDMs, and total modes in hybrid Kitaev ladder as a function of chemical potential $\mu$ at fixed $t_{\nu}=1.0$ and $\alpha<1.0$. For $0 \le \mu \le 0.5$, the system hosts two MZMs and two MDMs (total = 4). In the intermediate region $0.5 < \mu \le 1.5$, only two MDMs survive (total = 2), while for $\mu > 1.5$, all modes vanish, corresponding to a trivial phase.}
\label{EG_HKC_Modes}
\end{figure}

\par
To further elucidate the evolution of the total number of modes, we present a schematic variation of the number of MZMs, MDMs, and total modes as a function of the $\mu$ in fixed inter-leg coupling $t_{\nu}=1.0$ and the LR exponent $\alpha<1.0$, as shown in Fig.~\ref{EG_HKC_Modes}. In the regime $0 \le \mu \le 0.5$, the hybrid Kitaev ladder resides in a topological phase characterized by two MZMs with two MDMs, resulting in a total of four modes. As $\mu$ increases beyond $\mu=0.5$, the MDMs only appear. However, two MDMs persist in the intermediate region $0.5 < \mu \le 1.5$, indicating the presence of an LR exponent. For $\mu > 1.5$, no edge modes are present and the system becomes a trivial phase. A similar behaviour is observed for the total number of modes, which evolve through regions that host four, two, and no modes for $\alpha>1$. This behaviour clearly demonstrates how the chemical potential controls the topological edge modes as well as the persistence of bulk excitations in the crossover regime.

\section{Conclusion}
\label{conclusion}

In this work, we examined the topological properties of a hybrid Kitaev ladder composed of two parallel superconducting chains coupled through inter-leg coupling, where one leg hosts NN pairing, and the other incorporates LR pairing. By tuning the chemical potential $\mu$, the inter-leg coupling strength $t_{\nu}$ and the LR exponent $\alpha$, we demonstrated the emergence of a topological phase diagram~\cite{sw5x-pqpr, p79l-rty6}. Our analysis revealed multiple topological phases distinguished by the number of MZMs and MDMs. In particular, for $\alpha < 1$, the system supports a phase with two boundary localized MZMs and two MDMs, while for $\alpha > 1$, a transition to a phase hosting four MZMs occurs. This transition demonstrates the crucial role of LR pairing in reshaping the ladder's effective topological properties. In comparison with these regimes, we identified a crossover region characterized by the coexistence of MZMs and MDMs, reflecting hybridization between edge and bulk excitations. For sufficiently large values of $|\mu|$, the system becomes topologically trivial without edge modes. The observed reentrant structure further highlights the mutual influence between the chemical potential and the LR exponent. We also showed that the inter-leg coupling shifts the topological phase boundaries and modifies the stability region of the Majorana phases. The combined effect of $t_{\nu}$ and $\alpha$ therefore provides efficient control over the number and robustness of Majorana modes. In addition, the ladder geometry introduces a controllable topological proximity effect that enables manipulation of MZMs across coupled chains and offers a flexible framework for engineering Majorana hybridization and localization.
\par
Recent experimental progress suggests that such hybrid ladder systems are experimentally accessible. Engineered quantum-dot arrays permit site-resolved tuning of LR pairing and chemical potential, while topolectrical and superconducting circuit platforms allow controllable realization of effective hopping amplitudes and inter-leg couplings. These developments indicate that hybrid Kitaev ladders with tunable LR interactions can be implemented on realistic experimental platforms. Taken together, our results establish the hybrid Kitaev ladder as a promising platform for controllable Majorana manipulation and tunable topological superconductivity, with promising applications in topological quantum computation and quantum information technologies. In particular, the regime hosting four MZMs on the minimal ladder may provide an effective MZMs qubit subspace analogous to tetron architectures~\cite{qx36-4rv1, aghaee-2026}.
\par
Several directions can further extend the present study of the hybrid Kitaev ladder. such as time-dependent and periodically driven extensions of the hybrid Kitaev ladder offer a promising avenue. Periodic modulation of the pairing or inter-leg coupling could induce Floquet Majorana modes and dynamical topological phase, enabling further control over Majorana modes. Another interesting direction is the study of non-equilibrium dynamics, including quenches across topological phase boundaries and the associated dynamical phase transitions. This would allow one to probe the real-time formation and manipulation of edge modes in the presence of LR exponents. From an application perspective, exploring braiding protocols and quantum gate implementations within the ladder geometry would be highly relevant. The presence of multiple MZM sectors suggests the possibility of encoding a higher-dimensional topological qubit array. Investigating how inter-leg coupling can be used to controllably fuse and split Majorana modes is particularly important for quantum information processing. These directions collectively open the possibility of exploring richer topological phenomena and advancing the realization of controllable Majorana-based quantum devices.

\section*{Acknowledgement}
RK acknowledges  financial support from the Council of Scientific and Industrial Research (CSIR) fellowship, New Delhi, India [File No. 09/1217(15934)/2022-EMR-I].

\section*{DATA AVAILABILITY}
\label{DATA}
The data supporting the findings of this article are not publicly available. The data are available from the authors upon reasonable request.

%\section*{Subject Headings (PhySH)}
%Majorana bound states; Kitaev model; Topological phases of %matter; Topological phase transition; One-dimensional %systems; Tight-binding model, Topological Superconductors

\bibliography{Kitaev_Ladder}

\end{document}